%% file: journal.tex
\documentclass[journal]{IEEEtran}

\usepackage{amsthm}
\usepackage{float}
\usepackage{cite}										
\usepackage{amsmath}									
\usepackage{graphicx}
\usepackage{epstopdf}
\usepackage{float}
\usepackage{bm,upgreek}
\usepackage{amsfonts}
\usepackage{enumitem}
\usepackage{mathtools}
\usepackage{multirow}
\usepackage{booktabs}
\usepackage[subnum]{cases}
\usepackage{setspace}
\usepackage{color}
\usepackage{blkarray, bigstrut} 
\usepackage[caption=false,labelformat=simple]{subfig}

\usepackage{xcolor}
\usepackage{pgfplots}
\pgfplotsset{compat=1.5}
\usepackage{pgfplotstable}
\usepgfplotslibrary{statistics}
\pgfplotsset{grid style={dotted,gray}}
\usepackage{tikz}

\usepackage[toc,acronym]{glossaries}
\usepackage{mathtools,nccmath}
\usepackage[ruled]{algorithm}
\usepackage{algpseudocode,algorithmicx}

\algnewcommand{\LineComment}[1]{\State \(\#\) #1}

\allowdisplaybreaks

\ifCLASSINFOpdf
\else
\fi
\pgfplotsset{legend image with text/.style={legend image code/.code={%
\node[anchor=west, align=right] at (0.0cm,0cm) {#1};}},}

\makeatletter
\def\myline{\pgfutil@ifnextchar[{\my@line}{\my@line[]}}%
\def\my@line[#1](#2)(#3){%
\tikz[overlay] \draw[#1]  (#2)--(#3); 
}%
\algrenewcommand\algorithmicindent{1.0em}%

\newtheorem{example}{Example}

\makeatletter
\renewcommand{\ALG@beginalgorithmic}{\small}
\makeatother



\usepackage{etoolbox}
\BeforeBeginEnvironment{algorithmic}{\kern 0.3\baselineskip}
\AfterEndEnvironment{algorithmic}{\kern -0.1\baselineskip}

\newtheorem{definition}{Definition}

\makeatletter
\renewcommand{\ALG@beginalgorithmic}{\footnotesize}
\makeatother

\pgfplotsset{
    box plot/.style={
        /pgfplots/.cd,
        black,
        only marks,
        mark=-,
        mark size=\pgfkeysvalueof{/pgfplots/box plot width},
        /pgfplots/error bars/y dir=plus,
        /pgfplots/error bars/y explicit,
        /pgfplots/table/x index=\pgfkeysvalueof{/pgfplots/box plot x index},
    },
    box plot box/.style={
        /pgfplots/error bars/draw error bar/.code 2 args={%
            \draw  ##1 -- ++(\pgfkeysvalueof{/pgfplots/box plot width},0pt) |- ##2 -- ++(-\pgfkeysvalueof{/pgfplots/box plot width},0pt) |- ##1 -- cycle;
        },
        /pgfplots/table/.cd,
        y index=\pgfkeysvalueof{/pgfplots/box plot box top index},
        y error expr={
            \thisrowno{\pgfkeysvalueof{/pgfplots/box plot box bottom index}}
            - \thisrowno{\pgfkeysvalueof{/pgfplots/box plot box top index}}
        },
        /pgfplots/box plot
    },
    box plot top whisker/.style={
        /pgfplots/error bars/draw error bar/.code 2 args={%
            \pgfkeysgetvalue{/pgfplots/error bars/error mark}%
            {\pgfplotserrorbarsmark}%
            \pgfkeysgetvalue{/pgfplots/error bars/error mark options}%
            {\pgfplotserrorbarsmarkopts}%
            \path ##1 -- ##2;
        },
        /pgfplots/table/.cd,
        y index=\pgfkeysvalueof{/pgfplots/box plot whisker top index},
        y error expr={
            \thisrowno{\pgfkeysvalueof{/pgfplots/box plot box top index}}
            - \thisrowno{\pgfkeysvalueof{/pgfplots/box plot whisker top index}}
        },
        /pgfplots/box plot
    },
    box plot bottom whisker/.style={
        /pgfplots/error bars/draw error bar/.code 2 args={%
            \pgfkeysgetvalue{/pgfplots/error bars/error mark}%
            {\pgfplotserrorbarsmark}%
            \pgfkeysgetvalue{/pgfplots/error bars/error mark options}%
            {\pgfplotserrorbarsmarkopts}%
            \path ##1 -- ##2;
        },
        /pgfplots/table/.cd,
        y index=\pgfkeysvalueof{/pgfplots/box plot whisker bottom index},
        y error expr={
            \thisrowno{\pgfkeysvalueof{/pgfplots/box plot box bottom index}}
            - \thisrowno{\pgfkeysvalueof{/pgfplots/box plot whisker bottom index}}
        },
        /pgfplots/box plot
    },
    box plot median/.style={
        /pgfplots/box plot,
        /pgfplots/table/y index=\pgfkeysvalueof{/pgfplots/box plot median index},
        semithick,black
    },
    box plot width/.initial=1em,
    box plot x index/.initial=0,
    box plot median index/.initial=1,
    box plot box top index/.initial=2,
    box plot box bottom index/.initial=3,
    box plot whisker top index/.initial=4,
    box plot whisker bottom index/.initial=5,
}

\newcommand{\boxplot}[2][]{
    \addplot [box plot median,#1] table {#2};
    \addplot [forget plot, box plot box,#1] table {#2};
    \addplot [forget plot, box plot top whisker,#1] table {#2};
    \addplot [forget plot, box plot bottom whisker,#1] table {#2};
}

\algnewcommand\algorithmicforeach{\textbf{for each}}
\algdef{S}[FOR]{ForEach}[1]{\algorithmicforeach\ #1\ \algorithmicdo}

\include{acronyms}
\begin{document}

\newlist{myitemize}{itemize}{3}
\setlist[myitemize,1]{label=\textbullet,leftmargin=6.5mm}

\title{Observability Analysis for Large-Scale Power Systems Using Factor Graphs}

\author{Mirsad~Cosovic,~\IEEEmembership{Member,~IEEE,}
        Muhamed~Delalic,~\IEEEmembership{Student Member,~IEEE,}
        Darijo~Raca,~\IEEEmembership{Student Member,~IEEE,}
        Dejan Vukobratovic,~\IEEEmembership{Senior Member,~IEEE}

\thanks{M. Cosovic, M. Delalic and D. Raca are with Faculty of Electrical Engineering, University of Sarajevo, Bosnia and Herzegovina (e-mail: mcosovic@etf.unsa.ba, muha.delalic@gmail.com, draca@etf.unsa.ba), D. Vukobratovic is with Faculty of Technical Sciences, University of Novi Sad, Serbia, (email: dejanv@uns.ac.rs).}}

\markboth{}%
{Shell \MakeLowercase{\textit{et al.}}: Bare Demo of IEEEtran.cls for IEEE Journals}

\maketitle

\begin{abstract}
The state estimation algorithm estimates the values of the state variables based on the measurement model described as the system of equations. Prior to applying the state estimation algorithm, the existence and uniqueness of the solution of the underlying system of equations is determined through the observability analysis. If a unique solution does not exist, the observability analysis defines observable islands and further defines an additional set of equations (measurements) needed to determine a unique solution. For the first time, we utilise factor graphs and Gaussian belief propagation algorithm to define a novel observability analysis approach. The observable islands and placement of measurements to restore observability are identified by following the evolution of variances across the iterations of the Gaussian belief propagation algorithm over the factor graph. Due to sparsity of the underlying power network, the resulting method has the linear computational complexity (assuming a constant number of iterations) making it particularly suitable for solving large-scale systems. The method can be flexibly matched to distributed computational resources, allowing for determination of observable islands and observability restoration in a distributed fashion. Finally, we discuss performances of the proposed observability analysis using power systems whose size ranges between 1354 and 70000 buses.
\end{abstract}

\begin{IEEEkeywords}
Observability Analysis, Observable Islands, Observability Restoration, Belief Propagation, Factor Graphs, State Estimation, Large-Scale Systems
\end{IEEEkeywords}

\IEEEpeerreviewmaketitle

\section{Introduction}

\textbf{Motivation:} Deregulation of the electric power industry has led to the development of distributed control centres that operate over segments of a large interconnected power system. Such evolution of power systems calls for the distributed monitoring and control operations~\cite{poor}. As a part of the energy management system, power system monitoring relies on the distributed, scalable and computationally efficient state estimation (SE) routines that include the network topology processor, observability analysis, SE algorithm, and bad data processing.

Typically, the observability analysis enacts before the SE algorithm with bad data processing. According to the available measurements, observability analysis determines whether the network is observable or unobservable. Within an unobservable network, the observability analysis identifies all observable islands that can be solved independently, and obtains a minimal set of additional measurements providing the observability of the entire system, which ensures a unique estimate of the power system state~\cite{gou}. Over the past decades, comprehensive studies have been published that predominantly rely on \emph{distributed} SE algorithms with bad data processing. In contrast, \emph{computationally efficient} and \emph{distributed} observability analysis received significantly less attention from the research community. 

\textbf{Literature Review:} The identification of observable islands and observability restoration is traditionally carried out by applying numerical or topological methods. Numerical methods for the identification of observable islands and the observability restoration are mostly based on the matrix factorisation. For the identification of observable islands, there are several methods depending on the implementation approach: i) iterative methods that use the Jacobian~\cite{gou}, the gain~\cite{monticelliWu, wuMonticelli}, and the Gram matrix~\cite{garcia}; ii) non-iterative methods that use the gain matrix with the inverse of its triangular matrix~\cite{gouAbur}, and the Jacobian matrix based on the branch variable formulation~\cite{expositoAbur}. Similar categorisation can be made for the numerical methods applied to the observability restoration: i) iterative methods that use the gain~\cite{gouAbur, monticelliWu}, the Gram~\cite{garcia}, and reduced network matrix~\cite{contaxisKorres}; ii) non-iterative methods that use the Jacobian matrix~\cite{gou}, the gain matrix with the inverse of its triangular matrix~\cite{gouAbur2}, and the Gram matrix of the reduced Jacobian~\cite{korres2, korresManousakis}. These approaches depend on the accurate computation of zero pivots. The major drawback of these methods is the sensitivity to the zero pivot threshold value, where the incorrect value may deteriorate observability analysis. Also, non-iterative methods usually require definition and the calculation of additional matrices as compared to the iterative methods. Methods that use the Gram matrices, even for the reduced system, are prone to the risk of losing sparsity.

Unlike numerical methods, topological methods are less analysed with the most of the research efforts focused on the identification of the observable islands. These methods do not use any floating point calculations in the analysis. The decision is strictly based on the logical operations, and therefore requires information about the network connectivity, measurement types and their locations~\cite[Sec.~4.7]{abur}. Typically, topological methods first process all flow measurements to define initial islands, moving to processing injection measurements in an iterative fashion, followed by the identification of observable islands~\cite{horisberger, contaxisKorres, davis, clements}. In the paper \cite{slutsker}, authors propose a method based on the symbolic Jacobian matrix which considers only non-redundant measurements. Starting from a random bus, the proposed method utilise search technique to reveal observable islands through repetitive process.

Although topological methods can be adopted in a distributed framework, little work has been done in this area. However, numerical methods are also appealing for the observability analysis in the distributed framework, with many novel methods developed over the years. Several studies~\cite{korresDistributed, contaxisDistributed} proposed the distributed observability analysis based on the numerical methods that require coordination among areas, where observability restoration is accomplished by a non-iterative approach. Authors in~\cite{moura} analysed the distributed observability in terms of a consensus algorithm, providing conditions which enable solving the system of equations.

\textbf{Belief Propagation Approach and Contributions:} In this paper, we present a novel observability analysis approach based on a probabilistic graphical model, complementing our previous work on Gaussian belief propagation (GBP) based SE~\cite{CosovicTra}. The proposed method can be understood as both numerical and topological, with several advantages over the current state-of-the-art approaches. Our contributions are summarised as follows:
\begin{itemize}[leftmargin=*]
\item Compared to numerical methods, the proposed method does not utilise direct factorisation or inversion of matrices, avoiding inaccurate computation of zero pivots and incorrect choice of a zero threshold value.
\item Due to the sparsity of the underlying factor graph, the proposed method has \textit{linear} computational complexity per GBP iteration, making it particularly suitable for large-scale systems.
\item In the multi-area scenario, the proposed method can be implemented over the non-overlapping multi-area scenario without the central coordinator. The GBP algorithm ensures measurement data \textit{privacy} in the distributed architecture, where areas exchange only ``beliefs" about specific state variables. The GBP algorithm within local areas can be implemented using synchronous scheduling, where all messages in a given iteration are updated using the output of the previous iteration as an input. Message updates between local areas are not necessarily performed in full synchronism with the scheduling applied within local areas, as this would incur excessive communication delay to each iteration. Thus, the GBP framework allows the algorithm in which \textit{asynchronous} information exchange is possible~\cite{cosovicEUROCON}.
\item The method is straightforward to \emph{distribute} and \emph{parallelize}. In particular, even if implemented in the framework of centralised SE, it can be flexibly mapped to distributed computation resources (\eg parallel processing on graphical-processing units).
\end{itemize}
The novelty of our work follows from the fact that, for the first time, we adopt the GBP algorithm to carry out the observability analysis. Moreover, we emphasise a particularly simplifying feature of the proposed method, where within the GBP framework, it is sufficient to track the evolution of the variance values only. In particular, the GBP algorithm convergence depends on the convergence of the means, while variances always converge \cite{hanly}. To leverage this property, we propose observability analysis based only on variances resulting in an algorithm that \textit{always converges}. The resulting algorithm relies only on the graph structure and connections between nodes, and in contrast to numerical methods, it is immune to numerical issues.

\section{Observability Analysis}
\label{sec:oa}
This section provides a brief review of key concepts used for observability analysis in the power systems. Observability analysis is commonly performed on the linear decoupled measurement model~\cite[Ch.~7]{monticelliBook}, which we adopt in this paper. In summary, the observability analysis considers the following linearised measurement model:
\begin{equation}
    \mathbf{H}\mathbf{x} = \mathbf{z},
	\label{system_equations}
\end{equation}
where $\mathbf{z} \in \mathbb {R}^{m}$ is the vector of known values, $\mathbf{x}\in \mathbb {R}^{n}$ is the solution, while $\mathbf{H} \in \mathbb {R}^{m \times n}$ is a known matrix. The system is observable if and only if~\cite{gouAbur2}:
\begin{equation}
    \text{rank}(\mathbf{H})= n - 1,
	\label{rank}
\end{equation}
where the state variable corresponding to the slack bus is already included in \eqref{system_equations}. Ensuring the observability by condition \eqref{rank} does not imply achieving a good estimation of the state variables. This is because numerical ill-conditioning may deteriorate the SE algorithm, but in the most cases, if the system is observable, it can be efficiently estimated \cite{korres}. Note that the requirement posed by the condition~\eqref{rank} implicitly assumes using the least-squares approach in solving~\eqref{system_equations}. In theory, requirement~\eqref{rank} is not sufficient for the system~\eqref{system_equations} to have a solution (the existence of the solution depends on the vector $\mathbf{z}$). However,~\eqref{rank} guarantees invertibility of the symmetric matrix $\mathbf{H}^{{T}}\mathbf{H} \in \mathbb {R}^{n \times n}$, that is at the heart of the least-squares approach, therefore making~\eqref{system_equations} solvable.

When the given set of measurements is not sufficient (\ie $\text{rank}(\mathbf{H})<n - 1$), the observability analysis must identify all the possible observable islands that can be independently solved~\cite{gou}, where an observable island is defined as follows:
\begin{definition}
\label{def:1}
An observable island is a part of the power system for which the flows across all branches of the observable island can be calculated from the set of available measurements, independent of the values adopted for angular reference~\cite[Sec.~7.1.1]{monticelliBook}.
\end{definition}	
In other words, the observability analysis divides the system into the set of islands $\mathcal{S} = \{s_1,\dots,s_k\}$, where $k$ is the number of islands. Formally, the set of state variables $\mathcal{X} = \{x_1,\dots,x_n \}$ is divided into disjoint subsets $\mathcal{X}_i \subset \mathcal{X}$, $i=1,\dots,k$. Each island encloses $n_i = |\mathcal{X}_i|$ state variables, where the number of \emph{linearly independent} equations $l_i$ corresponding to the island $s_i$, is equal to $l_i = n_i-1$. Once the islands are determined, the observability analysis merges these islands in a way to protect previously-determined observable states from being altered by the new set of equations defined by the additional measurements. In general, this can be achieved by ensuring that the set of new measurements is a non-redundant set~\cite[Sec.~7.3.2]{monticelliBook}, \ie the set of equations must be linearly independent with regard to the global system. The aim of the observability restoration is to find this non-redundant set. The observability restoration finds exactly $k-1$ linearly independent equations, where the equation that defines the slack bus is known in advance.

\section{Belief Propagation Based Observability Analysis}
We consider a \emph{factor graph} approach for modelling the system of equations \cite{kschischang}. This approach starts with the construction of the factor graph with the set of \emph{variable nodes} representing unknown variables, and the set of \emph{factor nodes} corresponding to the set of equations. Note that a factor node connects to a variable node if and only if the unknown variable associated with the variable node is an argument of the corresponding equation associated with the factor node. Alongside the factor nodes that represent system equations (measurements), we introduce an additional set of virtual factor nodes defined as follows:
\begin{definition}
    The virtual factor node is a singly-connected factor node attached to the variable node, with the variance set\footnote{With a slight abuse of terminology, hereinafter we use the term ``set". However, one could argue that ``tend" could be deemed to be more appropriate.} either to zero or to infinity.
\end{definition}

\subsection{Determination of Observable Islands}
In our setup, from the set of equations, we retain a subset that corresponds to the set of the \emph{linearly independent} equations defined by the power flow and the power injection measurement functions. We start with the construction of the factor graph with the set of variable nodes $\mathcal{X}$ representing state variables, and the set of factor nodes $\mathcal{F}$ corresponding to the \emph{linearly independent} set of equations. Utilising the property that the GBP is robust to ill-conditioned scenarios caused by significant differences between measurement variances~\cite{CosovicTra, cosovicfast}, we augment the set of factor nodes $\mathcal{F}$ with the set of virtual factor nodes $\mathcal{F}_{\text{x}}$. Note that virtual factor nodes are equivalent to the pseudo-measurements of the voltage angles at the buses. Furthermore, if its variance tends to the infinity, the factor node has no influence on the rest of the graph.

\begin{example}[Constructing a factor graph] In this toy example, using 6-bus model presented in \figurename~\ref{fig:bus_branch}, we demonstrate the conversion from a bus/branch model with a given measurement configuration to the corresponding factor graph.
\begin{figure}[ht]
	\centering
	\includegraphics[width=5cm]{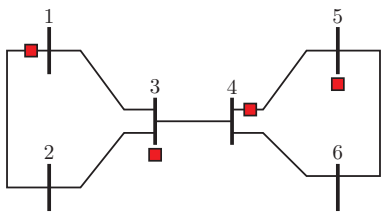}
	\caption{The 6-bus power network with a given measurement configuration.}
	\label{fig:bus_branch}
\end{figure} \noindent 
The linearised measurement Jacobian matrix can be formed as:
\begin{equation}
    \mathbf{H}=
    \begin{blockarray}{*{6}{c} l}
        \begin{block}{*{6}{>{$\footnotesize}c<{$}} l}
            $x_1$ & $x_2$ & $x_3$ & $x_4$ & $x_5$ & $x_6$ & \\
        \end{block}
        \begin{block}{[*{6}{r}]>{$\footnotesize}l<{$}}
            1 & -1 & 0 & 0 & 0 & 0 \bigstrut[t]& $M_{P_{12}}$ \\
            0 & 0 & 0 & 1 & -1 & 0 & $M_{P_{45}}$ \\
            -1 & -1 & 3 & -1 & 0 & 0 & $M_{P_{3}}$ \\
            0 & 0 & 0 & -1 & 2 & -1 & $M_{P_{5}}$ \\
        \end{block}
    \end{blockarray}
	\label{example1_Jacobian}
\end{equation}

The corresponding factor graph is given in \figurename~\ref{fig:detection_factorgraph}, where the set of variable nodes $\mathcal{X} = \{x_1,x_2,x_3,x_4,x_5,x_6\}$ is defined according to the vector of state variables $\mathbf{x}$. The set of factor nodes $\mathcal{F}$ is defined by corresponding measurements, where in our example, active power flows $M_{P_{12}}$, $M_{P_{45}}$, and active power injections $M_{P_{3}}$, $M_{P_{5}}$ measurements are mapped into factor nodes $\mathcal{F} = \{f_7,f_8,f_{9},f_{10}\}$ (red boxes), respectively. The factor graph reflects the structure of the matrix $\mathbf{H}$, where each row of the matrix corresponds to one factor node, while columns define variable nodes. Finally, we augment the factor graph with the set of virtual factor nodes $\mathcal{F}_\text{x} = \{f_{1},f_{2},f_{3},f_{4},f_{5},f_{6}\}$ (orange and green boxes), where the factor node $f_{1}$ is attached to the slack bus.
\begin{figure}[ht]
	\centering
	\includegraphics[width=5.1cm]{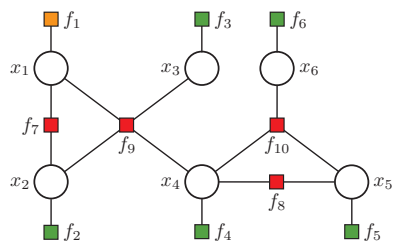}
	\caption{The factor graph that describes the structure of the Jacobian matrix~\eqref{example1_Jacobian} with additional set of virtual factor nodes.}
	\label{fig:detection_factorgraph}
\end{figure} 
\end{example}

Gaussian Belief-Propagation is a well-known algorithm for probabilistic inference on factor graphs \cite{loeliger}. It is used to efficiently calculate marginal distributions of random variables in a large-scale, sparse, linear systems of random variables corrupted by Gaussian noise. In this paper, we adapt GBP in order to solve the problem of observability analysis in power systems. We present GBP-based island detection in this subsection and extend it to GBP-based observability restoration in the next subsection.

The GBP-based island detection algorithm tracks the evolution of \emph{marginal} variances of the state variables, establishing observability criteria in the relation to whether or not these variances converge to a unique fixed point. Let us assume, for the time being, that the system is a fully observable, and the Jacobian matrix is defined according to power flows and injections measurement functions. According to Definition~\ref{def:1}, the values of the state variables corresponding to the buses are uniquely determined if at least one of the state variables in this set is declared known (\ie introducing voltage angle pseudo-measurement). As a result, the Jacobian matrix has a full rank \cite[Sec.~7.1.3]{monticelliBook}, while the marginal GBP variances converge to a unique fixed point. Indeed, according to Lemma 1 of~\cite{zhang} and Theorem 2 of~\cite{yates}, variances always converge to a unique fixed point if the fixed point exists. However, when the system is unobservable, using GBP-based approach it is still possible to identify subsets of state variables that converge to a unique point~\cite{yates}. We explore this feature for the identification of observable islands.

We formalise the proposed setup used to detect observable islands in Algorithm~\ref{alg1}. We consider a factor graph with the set of variable nodes $\mathcal{X}$, the set of factor nodes $\mathcal{F}$, and the set of virtual factor nodes $\mathcal{F}_{\text{x}}$ constructed as follows. The set $\mathcal{F}_{\text{x}}$ contains $n$ factor nodes, out of which a single one has a variance set to zero and is attached to an arbitrary (randomly selected) variable node (we call this factor node a ``probe"), and the remaining $n-1$ factor nodes have variances set to infinity, and are attached to the remaining $n-1$ variable nodes. After the initialisation (lines 1-9), the identification of observable islands procedure (lines 10-40) starts by selecting a random variable node from the set $\mathcal{X}^{(\nu)}$ to which a probe factor node is attached, where $\nu$ denotes the number of passes through the procedure (line 12). The message-passing starts by setting initial GBP variances from variable nodes $x_i \in \mathcal{X}^{(\nu)}$ and factor nodes $f_j \in \mathcal{F}^{(\nu)}$ to finite positive values (line 13). The message-passing framework with marginal inference (lines 14-28) represents the \textit{main} algorithm routine which determines the observable islands (lines 29-34), where the set $\mathcal{F}_i \setminus f_j$ represents factor nodes incident to the variable node $x_i$, excluding the factor node $f_j$, while the set $\mathcal{X}_j\setminus x_i$ represents variable nodes incident to the factor node $f_j$, excluding the variable node $x_i$. Upon the calculation of the marginal variances for all the variable nodes in $\mathcal{X}^{(\nu)}$, a node $x_i \in \mathcal{X}^{(\nu)}$ is proclaimed observable, \ie $x_i \in \mathcal{X}_{\text{o}}^{(\nu)}$, if and only if its marginal variance $\hat v_{x_i}$ is finite. Note that the resulting observable island depends on the position of the probe factor node. After the discovery of observable variable nodes, we consider a subgraph of the factor graph that contains the set of unobservable variable nodes $\mathcal{X}^{(\nu+1)} =$ $\mathcal{X}^{(\nu)} \setminus$ $ \mathcal{X}_\text{o}^{(\nu)}$ and the set of all factor nodes $\mathcal{F}^{(\nu+1)}$ and $\mathcal{F}_\text{x}^{(\nu+1)}$ connected exclusively to variable nodes in $\mathcal{X}^{(\nu+1)}$ (lines 35-37). The procedure now repeats from the line 11. The next observable island is revealed by introducing a new randomly selected probe, and setting its virtual factor node variance to zero. This causes a change in marginal variances, in a way that all unobservable nodes within the next revealed observable island switch over into finite marginal variance values. The procedure is repeated until all observable islands are revealed.

\begin{algorithm}[ht]
\caption{Identification of Observable Islands}
\label{alg1}
\begin{spacing}{1.25}
\begin{algorithmic}[1]
\Procedure {Initialisation}{}
  \State {determine linearly independent set of equations}
  \ForEach{$x_i \in \mathcal{X}$}
  	\State assign virtual factor node $f_i \in \mathcal{F}_{\text{x}}$ with infinite variance 
  \EndFor 
  \State initialise counter $\nu = 1$
  \State initialise the set of variable nodes $\mathcal{X}^{(\nu)} = \mathcal{X}$ 
  \State initialise the set of factor nodes $\mathcal{F}_{\text{x}}^{(\nu)} = \mathcal{F}_{\text{x}},$ $\mathcal{F}^{(\nu)} = \mathcal{F}$
\EndProcedure
\myline[black, line width=0.15mm](-1.7,3)(-1.7,0.25)

\Procedure {Identification of Observable Islands}{}
  \While {$\mathcal{X}^{(\nu)} \neq \emptyset$}
	\State {set a random factor node $f_{p} \in \mathcal{F}_{\text{x}}^{(\nu)}$ to zero variance (probe)}
	\State {set all variances from $x_i \in \mathcal{X}^{(\nu)}$ to $f_j \in \mathcal{F}^{(\nu)}$ on finite values}

	\Procedure {Message-Passing Framework}{}
	  \While {stopping criterion is not met}
		\ForEach{$f_j \in \mathcal{F}^{(\nu)}$}
			\State\vspace*{-1.1\baselineskip}
        	\begin{fleqn}[\dimexpr\leftmargini--\labelsep]
        	\setlength\belowdisplayskip{0pt}
			\begin{equation*}
        	\begin{aligned}
        	\;\;\;\;\;\;\;\;\;v_{f_j \to x_i}^{(\tau)} = 
			\sum_{x_b \in \mathcal{X}_j \setminus x_i} 
			v_{x_b \to f_j}^{(\tau-1)}
        	\end{aligned}
        	\nonumber
			\label{SE_Gauss_mth}
			\end{equation*}
			\end{fleqn}%
		\EndFor

		\ForEach{$x_i \in \mathcal{X}^{(\nu)}$}
			\State\vspace*{-1.1\baselineskip}
        	\begin{fleqn}[\dimexpr\leftmargini--\labelsep]
        	\setlength\belowdisplayskip{0pt}
			\begin{equation*}
      	 	\begin{aligned}
        	\;\;\;\;\;\;\;\;\;\left({v_{x_i \to f_{j}}^{(\tau)}}\right)^{-1} = 
			\sum_{f_a \in \mathcal{F}_i\setminus f_j} 
			\left({v_{f_{a} \to x_i}^{(\tau-1)}}\right)^{-1}
        	\end{aligned}
        	\nonumber
			\label{SE_Gauss_mth}
			\end{equation*}
		    \end{fleqn}%
		\EndFor
	  \EndWhile 
	\EndProcedure 
    \myline[black, line width=0.15mm](-1.7,4.37)(-1.7,0.25)

	\Procedure {Marginal Variances}{}
		\ForEach {$x_i \in \mathcal{X}^{(\nu)}$}
     		\State\vspace*{-1.3\baselineskip}
        	\begin{fleqn}[\dimexpr\leftmargini--\labelsep]
        	\setlength\belowdisplayskip{0pt}
			\begin{equation*}
        	\begin{gathered}
			\;\;\;\;\;\left({\hat v_{x_i}}\right)^{-1} = 
			\sum_{f_a \in \mathcal{F}_i} 
			\left({v_{f_{a} \to x_i}^{(\tau_{\max})}}\right)^{-1}
			\end{gathered}
			\label{marginal}
			\end{equation*}
        	\end{fleqn}%
		\EndFor
	\EndProcedure
    \myline[black, line width=0.15mm](-1.7,1.82)(-1.7,0.25)
	\State initialise the set of observable variable nodes 
	$\mathcal{X}_\text{o}^{(\nu)} = \emptyset$
	\ForEach{$x_i \in \mathcal{X}^{(\nu)}$}
  		\If {$\hat v_{x_i}$ has finite variance}
  			\State the variable node is observable $\mathcal{X}_\text{o}^{(\nu)}\cup x_i$
		\EndIf
	\EndFor 

	\State {the set $\mathcal{X}_\text{o}^{(\nu)}$ form observable island $s_\nu$} 
	\State {form $\mathcal{X}^{(\nu+1)} = \mathcal{X}^{(\nu)} \setminus \mathcal{X}_\text{o}^{(\nu)}$} 
	\State {form $\mathcal{F}^{(\nu+1)},$ $\mathcal{F}_\text{x}^{(\nu+1)}$ incident only to variable nodes $\mathcal{X}^{(\nu+1)}$}
	\State increase counter $\nu := \nu + 1$
  \EndWhile 
\EndProcedure
\myline[black, line width=0.15mm](-1.7,13.05)(-1.7,0.25)
\end{algorithmic}
\end{spacing}
\end{algorithm}
\setlength{\textfloatsep}{15pt}

\begin{example}[Identification of Observable Islands]\label{example:ident} We demonstrate the GBP-based island detection algorithm, using the factor graph given in Example 1. We start by selecting the probe factor node $f_1$ and setting its variance to zero, while variances of the rest of virtual factor nodes are set to infinity:
\begin{equation}
	\begin{aligned}
        v_{f_1 \to x_1} &\to 0 \\
        v_{f_i \to x_i} &\to \infty, \;\;\; i=2,\dots,6.
    \end{aligned}
\end{equation}

The algorithm starts with initial finite (\eg equal to one) variance values:
\begin{equation}
    v_{x_i \to f_i}^{(0)} = 1, \;\;\; x_i \in \mathcal{X},\; f_i \in \mathcal{F}.
\end{equation}
The GBP iteration $\tau = 1$ starts with computing variances from factor nodes $\mathcal{F}$ to variable nodes $\mathcal{X}$, using incoming variances from variable nodes $\mathcal{X}$ to factor nodes $\mathcal{F}$. For example, computation of the variance from the factor node $f_{9}$ to the variable node $x_1$ is computed as follows:
\begin{equation}
    v_{f_{9} \to x_1}^{(1)} = v_{x_2 \to f_{9}}^{(0)} + v_{x_3 \to f_{9}}^{(0)} + v_{x_4 \to f_{9}}^{(0)} = 3.
\end{equation}
Next, the algorithm proceeds with computing variances from variable nodes $\mathcal{X}$ to factor nodes $\mathcal{F}$, using incoming variances from factor nodes $\mathcal{F} \cup \mathcal{F}_\text{x}$ to variable nodes $\mathcal{X}$, for example:
\begin{equation}
    \cfrac{1}{v_{x_1 \to f_7}^{(1)}} = \cfrac{1}{v_{f_{9} \to x_1}^{(1)}} +\cfrac{1}{v_{f_1 \to x_1}}; \;\; \;\; v_{x_1 \to f_7}^{(1)} \to 0.
\end{equation}

When the algorithm converges, all variances from factor nodes $\mathcal{F}$ to variable nodes $\mathcal{X}$ are equal to infinity, except the variance from the factor node $f_7$ to the variable node $x_2$ that tends to zero $v_{f_7 \to x_2}^{(\tau_{\max})} \to 0$. The marginal variance of variable nodes $\mathcal{X}$ can be obtained using variances from factor nodes $\mathcal{F} \cup \mathcal{F}_\text{x}$ to variable nodes $\mathcal{X}$:
\begin{equation}
	\begin{aligned}
        \cfrac{1}{\hat v_{x_1}} = \cfrac{1}{v_{f_7 \to x_1}^{(\tau_{\max})}} + \cfrac{1}{v_{f_{9} \to x_1}^{(\tau_{\max})}} +\cfrac{1}{v_{f_1 \to x_1}}; \;\; \;\; \hat v_{x_1} \to 0 \\
    \end{aligned}
\end{equation}
\begin{equation}
	\begin{aligned}
        \cfrac{1}{\hat v_{x_2}} = \cfrac{1}{v_{f_7 \to x_2}^{(\tau_{\max})}} + \cfrac{1}{v_{f_{9} \to x_2}^{(\tau_{\max})}} +\cfrac{1}{v_{f_2 \to x_2}}; \;\; \;\; \hat v_{x_2} \to 0,
    \end{aligned}
\end{equation}
while other marginals are equal to infinity. Hence, the first island $s_1$ is defined by state variables $\mathcal{X}_1 = \{x_1,x_2\}$.

Next, we observe the modified network and the measurement configuration shown in Fig. \ref{fig:detection_factorgraph1}, where the new probe factor node is $f_4$. We repeat the process revealing the second island $s_2$ defined by state variables $\mathcal{X}_2 = \{x_4,x_5,x_6\}$. Finally, the third island $s_3$ encapsulates the state variable $\mathcal{X}_3 = \{x_3\}$.
\begin{figure}[ht]
	\centering
	\includegraphics[width=3.2cm]{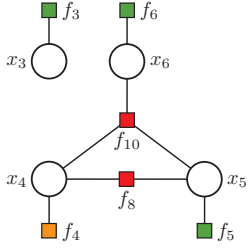}
	\caption{The modified factor graph after the first island is detected.}
	\label{fig:detection_factorgraph1}
	\end{figure} 
\end{example}

Astonishingly, the GBP-based method for the identification of observable islands is similar to the decoding of low-density parity-check codes over erasure channels \cite{spielman}, which may inspire further work in this domain.

\subsection{Observability Restoration}
For the observability restoration, we consider the unobservable system divided into $k$ islands, where each island has $l_i = n_i - 1$ linearly independent equations. Putting the set of linear equations from all $k$ observable islands together, we have: 
\begin{equation}
    \mathbf{H}_{\mathrm{u}}\mathbf{x} = \mathbf{z}_{\mathrm{u}},
	\label{system_equations_ind}
\end{equation}
where $\mathbf{H}_{\mathrm{u}} \in \mathbb{R}^{h \times n}$, $\mathbf{x} \in \mathbb{R}^{n}$, $\mathbf{z}_{\mathrm{u}} \in \mathbb{R}^{h}$, and $h=n-k$. In addition, we introduce a new set of equations containing all the candidate measurements that are able to merge islands:
\begin{equation}
    \begin{bmatrix} \mathbf{H}_{\text{c}} \\ \mathbf{H}_{\text{p}} \end{bmatrix} 
    \mathbf{x} = 
    \begin{bmatrix} \mathbf{z}_{\text{c}} \\ \mathbf{z}_{\text{p}} \end{bmatrix},
	\label{system_equations_candidate}
\end{equation}
where $\mathbf{H}_{\mathrm{c}} \in \mathbb{R}^{q \times n}$ and $\mathbf{z}_{\mathrm{c}} \in \mathbb{R}^{q}$ represent measurements which were rejected during observable island identification, while $\mathbf{H}_{\mathrm{p}} \in \mathbb{R}^{b \times n}$ and $\mathbf{z}_{\mathrm{p}} \in \mathbb{R}^{b}$ describe pseudo-measurements. 
The system of equations described with $\mathbf{H}_{\mathrm{c}}$ contains only linearly independent equations. Furthermore, we note that additional $w = k - q - 1$ equations among equations defined by pseudo-measurements are needed to achieve observability, as the equation that defines the slack bus is known in advance.

Let us introduce the vectors $\boldsymbol {\upalpha}$, $\boldsymbol {\upbeta}$ and $\boldsymbol {\upgamma}$ representing an uncorrelated noise with the independent and identically distributed entries, where each element follows a zero-mean Gaussian distribution with associated variances $\alpha_i$, $\beta_i$ and $\gamma_i$ Hence, systems~\eqref{system_equations_ind} and~\eqref{system_equations_candidate} can be merged together as: 
\begin{equation}
    \begin{bmatrix} 
            \mathbf{H}_{\text{u}} \\ \mathbf{H}_{\text{c}} \\ \mathbf{H}_{\text{p}} 
    \end{bmatrix} 
    \mathbf{x} + 
    \begin{bmatrix} 
        \boldsymbol {\upalpha} \\ \boldsymbol {\upbeta} \\  \boldsymbol {\upgamma}  
    \end{bmatrix} = 
    \begin{bmatrix} 
        \mathbf{z}_{\text{u}} \\ \mathbf{z}_{\text{c}} \\ \mathbf{z}_{\text{p}} 
    \end{bmatrix}.
	\label{system_equations_aug}
\end{equation}
Next, we apply a simple trick using the values of variances assigned to different measurements. Namely, we first set $\alpha_i \to 0$, $i=1,\dots,h$, $\beta_i \to 0$, $i=1,\dots,q$, and $\gamma_i \to \infty$, $i=1,\dots,b$. Then, we select exactly $w$ linearly independent equations (with regards to the global system) from the set of pseudo-measurements. Without loss of generality, we choose the first $w$ equations and we set $\gamma_i \to 0$, $i=1,\dots,w$, \ie we change the variance of $w$ linearly independent measurements from infinity to zero. After additional $w$ equations are ``denoised", the solution of \eqref{system_equations_aug} will be determined only by the equations associated to the variances that are set to zero. Note that the corresponding \emph{residuals} are also equal to \textit{zero}. In other words, each of the noiseless equations is a critical equation~\cite[Sec.~5.2]{abur}. However, the described scenario cannot be solved using the traditional weighted least-squares method, because we defined the ill-conditioned scenario caused by significant differences between variances. In contrast, the GBP approach is robust to ill-conditioned scenarios, and above-mentioned scenario can be solved~\cite{cosovicfast}.

Introducing the additional vector of measurement residuals $\mathbf{r} \in \mathbb{R}^{d}$, $d = h + q + b$, we can expand \eqref{system_equations_aug} as follows:
\begin{equation}
    \begin{bmatrix} 
        \begin{bmatrix} 
            \mathbf{H}_{\mathrm{u}} \\ \mathbf{H}_{\text{c}} \\ \mathbf{H}_{\text{p}} 
        \end{bmatrix} & 
        \mathbf{I}_{d \times d} \\[8pt]
        \mathbf{0}_{d \times n} & \mathbf{I}_{d \times d}
    \end{bmatrix} 
    \begin{bmatrix} 
        \mathbf{x} \\ \mathbf{r} 
    \end{bmatrix}  + 
    \begin{bmatrix} 
       \boldsymbol {\upalpha} \\ \boldsymbol {\upbeta} \\  \boldsymbol {\upgamma}  \\ \boldsymbol {\updelta}  
    \end{bmatrix} = 
    \begin{bmatrix} 
        \mathbf{z}_{\text{u}} \\ \mathbf{z}_{\text{c}} \\ \mathbf{z}_{\text{p}} \\ \mathbf{0}_{d\times 1} 
    \end{bmatrix},
	\label{system_equations_fg}
\end{equation}
where $\mathbf{I}$ is the identity matrix, $\mathbf{0}_{d \times n}$ is the all-zero matrix, $\mathbf{0}_{d \times 1}$ is the vector of zeros, and the vector $\boldsymbol {\updelta}$ represents an uncorrelated noise defined in the same way as vectors $\boldsymbol {\upalpha}$, $\boldsymbol {\upbeta}$ and $\boldsymbol {\upgamma}$. By applying the weighted least-squares method to \eqref{system_equations_fg} and inspecting the variance-covariance matrix, it is easy to show that the variances of the state variables $\mathbf{r}$ depend only on the variances related to the noise vector $\boldsymbol {\updelta}$ in the case when the system \eqref{system_equations_aug} contains the set of linearly independent equations, and therefore \eqref{system_equations_aug} does not influence the residual values. In other words, if an equation described by a factor node $f_i$, sends infinity variance to the corresponding residual described by a variable node $r_i$, the equation is linearly independent. 

\begin{figure}[ht]
	\centering
	\includegraphics[width=5.7cm]{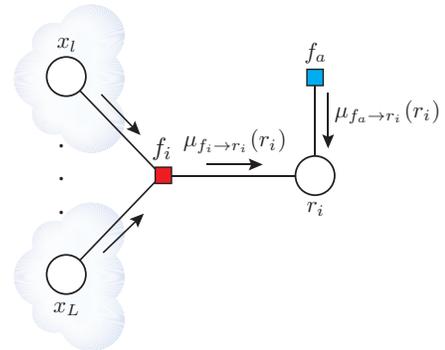}
	\caption{A part of the factor graph that corresponds to a single equation.}
	\label{fig:fgpart}
\end{figure} 

For example, consider a part of the factor graph shown in \figurename~\ref{fig:fgpart}, and focus on a single measurement that defines the factor node $f_i$. Note that the message from a factor node to a variable node represents a ``belief" about the variable node, and each message exchanged in GBP is completely represented using only two values: mean and variance. Messages from variable nodes $\{x_l,\dots,x_L\}$ to the factor node $f_i$ carry a collective evidence of the rest of the factor graph, and according to these messages, the GBP algorithm computes the message $\mu_{f_i \to r_i}(r_i)$ from the factor node $f_i$ to the variable node $r_i$. Next, we set the mean value of the message $\mu_{f_a \to r_i}(r_i)$ from the singly-connected factor node $f_a$ to the variable node $r_i$ to zero, and its variance to a finite positive value. The marginal distribution of the variable node $r_i$ is obtained as the product of the incoming messages $\mu_{f_i \to r_i}(r_i)$ and $\mu_{f_a \to r_i}(r_i)$. The mean of the marginal will remain zero only if the variance of the message $\mu_{f_i \to r_i}(r_i)$ tends to infinity. When this is the case, the corresponding equation is linearly independent and does not affect the residual value. 

To summarize, the GBP-based algorithm decomposes the contribution of each factor node (equation) to a group of variable nodes (state variables and a measurement residual). This decomposition allows insight into the structure of the measurement residual, where the impact of each measurement can be observed. Hence, if one factor node $f_i$ sends finite variance value to the residual $r_i$, this means that the equation has an impact on the measurement residual value, \ie $r_i \neq 0$, while infinite variance corresponds to the case when the equation does not have impact to the measurement residual, \ie $r_i = 0$.

In order to reduce numerical complexity, it is possible to avoid running GBP inference across the entire factor graph defined by \eqref{system_equations_fg}. Instead, we can transform the original factor graph into a considerably more compact version which is equivalent from the restoration viewpoint. The variable nodes from sets $\mathcal{X}_i$, $i = 1, \dots, k$ are connected by factor nodes $\mathcal{F}_\mathrm{c}$ and $\mathcal{F}_\mathrm{p}$ defined according to matrices $\mathbf{H}_{\mathrm{c}}$ and $\mathbf{H}_{\mathrm{p}}$, respectively. Also, each island $s_i$ has $l_i = n_i-1$ linearly independent equations, where revealing any of the state variables from the set $\mathcal{X}_i$ reveals the entire observable island $s_i$. This allows the construction of the factor graph, where all state variables $\mathcal{X}_i$ enclosed by the island $s_i$ are merged into a single variable node \ie each island becomes the variable node. Thus, we observe the factor graph with a set of variable nodes $\mathcal{S} = \{s_1,\dots,s_k\}$ and the set of factor nodes $\mathcal{F}_\mathrm{c} \cup \mathcal{F}_\mathrm{p}$. Same as before, we introduce the set of virtual factor nods $\mathcal{F}_{\text s}$ attached to the variable nodes from the set $\mathcal{S}$. This strategy reveals an exciting finding: the factor graph defined by the variable nodes $\mathcal{S}$, and factor nodes $\mathcal{F}_\mathrm{c}$ and $\mathcal{F}_\mathrm{p}$ equates to the matrix $\mathbf{W}_{\text {bc}}$ given in \cite{korres2}. However, in contrast to \cite{korres2} where the Gram matrix is needed, we are solving the observability restoration problem directly over the matrix $\mathbf{W}_{\text {bc}}$ using the GBP algorithm.

\begin{algorithm}[ht]
\caption{Observability Restoration}
\label{alg2}
\begin{spacing}{1.25}
\begin{algorithmic}[1]

\Procedure {Initialisation}{}
    \ForEach{$f_i \in \mathcal{F}_\text{c}$}
        \State assign finite variance equal to $v_i$
    \EndFor
    \ForEach{$f_i \in \mathcal{F}_\text{p}$}
        \State assign infinite variance
    \EndFor
     \ForEach{$s_i \in \mathcal{S}$}
        \State assign virtual factor node $f_i \in {\mathcal{F}}_{\text s}$ with infinite variance 
    \EndFor
    \State initialise $w = k - q - 1$ 
    \State initialise counter $\nu = 1$
\EndProcedure
\myline[black, line width=0.15mm](-1.7,4.6)(-1.7,0.25)

\Procedure {Observability Restoration}{}
    \While {$w > 0$}
	    \State {select $\nu$-th factor node $f_\nu \in \mathcal{F}_\text{p}$ and assign variance $v_i$}
	    \State {find $s_\nu$ as one of the variable nodes incident to the factor node $f_\nu$}
	    \State {assign zero variance to the virtual factor node attached to $s_\nu$}
	   	\State {set all variances from $s_i \in {\mathcal{S}}$ to $f_j \in \mathcal{F}_\text{c} \cup \mathcal{F}_\text{p}$ on finite values}
	    \Procedure {Message-Passing Framework}{}
	        \While {stopping criterion is not met}
		        \ForEach{$f_j \in \mathcal{F}_\text{c} \cup \mathcal{F}_\text{p}$ }
			        \State\vspace*{-1.1\baselineskip}
        	        \begin{fleqn}[\dimexpr\leftmargini--\labelsep]
        	        \setlength\belowdisplayskip{0pt}
			        \begin{equation*}
        	        \begin{aligned}
        	        \;\;\;\;\;\;\;\;\;v_{f_j \to s_i}^{(\tau)} = v_j + \sum_{s_b \in {\mathcal{S}}_j \setminus s_i} 
		        	v_{s_b \to f_j}^{(\tau-1)}
        	        \end{aligned}
        	        \nonumber
			        \end{equation*}
			        \end{fleqn}%
		        \EndFor

		        \ForEach{$s_i \in {\mathcal{S}} $}
			        \State\vspace*{-1.1\baselineskip}
        	        \begin{fleqn}[\dimexpr\leftmargini--\labelsep]
        	        \setlength\belowdisplayskip{0pt}
			        \begin{equation*}
      	 	        \begin{aligned}
        	        \;\;\;\;\;\;\;\;\;\left({v_{s_i \to f_{j}}^{(\tau)}}\right)^{-1} = 
			        \sum_{f_a \in \mathcal{F}_i\setminus f_j} \left({v_{f_{a} \to s_i}^{(\tau-1)}}\right)^{-1}
        	        \end{aligned}
        	        \nonumber
			        \label{SE_Gauss_mth}
			        \end{equation*}
		            \end{fleqn}%
		        \EndFor
	        \EndWhile 
	    \EndProcedure
        \myline[black, line width=0.15mm](-1.7,4.37)(-1.7,0.25)
        
        \Procedure {Linear Independence}{} 
            \State {assign residual variable node $r_\nu$ to the factor node $f_\nu$}
  		    \State {compute $v_{f_\nu \to r_\nu}$}
  		    \If {$v_{f_\nu \to r_\nu} \to \infty$}
  			   \State factor node $f_\nu$ represents independent equation 
  			   \State $w := w - 1$ 
  			\Else
  			    \State factor node $f_\nu$ represents dependent equation
  			    \State assign infinite variance to $f_\nu$
		   \EndIf
		   \State change variance of the virtual factor node of $s_\nu$ to infinity
		    \State increase counter $\nu := \nu + 1$
	    \EndProcedure 
        \myline[black, line width=0.15mm](-1.7,4.6)(-1.7,0.25)
    \EndWhile 
\EndProcedure
\myline[black, line width=0.15mm](-1.7,12.67)(-1.7,0.25)
\end{algorithmic}
\end{spacing}
\end{algorithm} 
\setlength{\textfloatsep}{15pt}

We formalise the proposed method for the observability restoration in Algorithm~\ref{alg2}. The initialisation step (lines 1-13) creates the factor graph, where each factor node $f_i$ from the set $\mathcal{F}_\mathrm{c}$ has the same variance equal to $v_i$, \ie variance $v_i$ corresponds to the equation associated with the factor node $f_i$. Each variable node $s_i \in \mathcal{S}$ has the virtual factor node with variances set to infinity (lines 8-10). The algorithm continues observability restoration procedure (lines 14-44) until $w = k - q - 1$ linearly independent equations are found (line 11). The observability restoration procedure (lines 14-44) starts by selecting a factor node $f_\nu$ from the set $\mathcal{F}_\mathrm{p}$ followed by changing its variance from the infinity to $v_i$. In other words, we select one of the equations defined by pseudo-measurements that can merge islands (line 16). Then, we select one of the variable nodes incident to the factor node $f_\nu$ and change the variance of its virtual factor node to zero. This virtual factor node has a role of a slack variable node that exists in the system (lines 17-18). The message-passing procedure (lines 20-29) commence after the initial GBP variances from variable nodes $s_i \in {\mathcal{S}}$ to factor nodes $f_j \in \mathcal{F}_\text{c} \cup \mathcal{F}_\text{p}$ are set to finite positive values (line 19), where the set $\mathcal{F}_i \setminus f_j$ represents factor nodes incident to the variable node $s_i$, excluding the factor node $f_j$, while the set ${\mathcal{S}}_j\setminus s_i$ represents variable nodes incident to the factor node $f_j$, excluding the variable node $s_i$. The algorithm proceeds with the linear independence procedure (lines 30-42). The equation, described by the factor node $f_\nu$, is linearly dependent or independent, based on the variance value from factor node $f_\nu$ to the residual variable node $r_\nu$. The linear independence procedure marks an equation as linearly independent if the variance $v_{f_\nu \to r_\nu}$ has an infinite value (line 34). Thus, the number of required equations to provide observability is reduced by one (line 35). In contrast, the linearly dependent equation is excluded from the graph by changing its variance from $v_i$ to infinity (lines 37-38), and the algorithm proceeds by selecting a new factor node (line 16). The procedure is repeated until observability of the entire system is achieved.

\begin{example}[Observability Restoration] We demonstrate the GBP-based observability restoration method by leveraging Example~\ref{example:ident}, where we identified 3 islands $\mathcal{S} = \{s_1,s_2,s_3\}$, as depicted in \figurename~\ref{fig:islands}.
\begin{figure}[!ht]
	\centering
	\includegraphics[width=5.45cm]{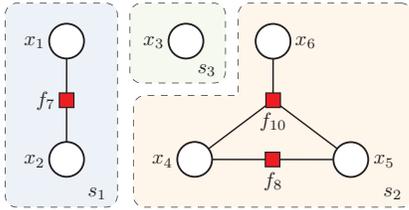}
	\caption{Observable islands of the 6-bus power network.}
	\label{fig:islands}
\end{figure}

We assume pseudo-measurements can merge islands only if injected at the boundary buses of observable islands. Thus, we consider pseudo-measurements $M_{P_1}$, $M_{P_2}$, and $M_{P_4}$, which define the set $\mathcal{F}_\text{p} = \{ f_{12}, f_{13}, f_{14}\}$, respectively. The injection measurement $M_{P_3}$ which was rejected during observable island identification define the set $\mathcal{F}_{\text c} = \{f_{11}\}$ (\figurename~\ref{fig:restoration}, red box), with finite positive value variance (\eg equal to one). This measurement represents a linearly independent equation and it is added at the beginning by default. 
\begin{figure}[ht]
	\centering
	\includegraphics[width=3.5cm]{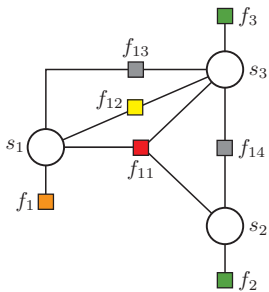}
	\caption{The factor graph with pseudo-measurements.}
	\label{fig:restoration}
\end{figure}

The resulting factor graph is shown in \figurename~\ref{fig:restoration}, where we first select the factor node $f_{12}$ (yellow box) and set its variance to a finite positive value, while the variances of factor nodes $f_{13}$ and $f_{14}$ (grey boxes) are set to infinity. We begin by setting variances from the virtual factor nodes as follows: 
\begin{equation}
	\begin{aligned}
        v_{f_1 \to s_1} & \to 0 \\
        v_{f_i \to s_i} &\to \infty, \;\;\; i=2,3.
    \end{aligned}
\end{equation}

The GBP iterations are performed in a similar way as in Example~\ref{example:ident}, except that here we use variance $v_i$ to compute variances from factor nodes $f_i \in \mathcal{F}_\text{c} \cup \mathcal{F}_\text{p}$ to variable nodes $s_i \in {\mathcal{S}}$. When the algorithm converges, we include the residual variable node $r_{12}$ incident to the factor node $f_{12}$ and compute variance $v_{f_{12} \to r_{12}}$ using variances into the factor node $f_{12}$:
	\begin{equation}
        v_{f_{12} \to r_{12}}^{(\tau_{\max})} = v_{s_1 \to f_{12}}^{(\tau_{\max})} + v_{s_3 \to f_{12}}^{(\tau_{\max})}; \;\; \;\; v_{f_{12} \to r_{12}}^{(\tau_{\max})} \to \infty.
	\end{equation}
Thus, the factor node $f_{12}$ represents a linearly independent equation, and the system is fully observable using measurements $M_{P_1}$ and $M_{P_3}$ described by factor nodes $f_{11}$ and $f_{12}$. respectively. 
\end{example}

\section{Numerical Results}
In this section, we assess our proposed method against the numerical and topological state-of-the-art methods available in the literature. In our evaluation, we explore a large set of power systems with different bus configurations. We evaluate the performance of the proposed method using power systems with 1354, 2000, 10000, 25000 and 70000 buses \cite{wehenkel, birchfield}. For each power system, we generate 100000 \emph{random} measurement configurations in order to get statistically significant results. The measurement set includes active power flow on both branch sides and active power injection measurements. For each power system size, we use different redundancies to control for the resulting number of observable islands, where we aim at producing approximately the same number of measurement configurations per each number of islands ranging between 2 and 33.

In all scenarios, we use box plots where we normalise execution time of each of the different state-of-the-art methods $t_{\text{sa}}$ by the execution time of the GBP-based approach $t_{\text{bp}}$. Hence, the case $t_{\text{sa}} / t_{\text{bp}} > 1$ corresponds to the case where the GBP-based method shows better performance (\ie lower execution time), while $t_{\text{sa}} / t_{\text{bp}} < 1$ corresponds to the case where the GBP-based method shows worse performance (\ie higher execution time) in comparison with the state-of-the-art method. The methods have been implemented using Julia programming language, while the source code of the GBP-based observability analysis can be found as the part of the JuliaGrid package\footnote{https://github.com/mcosovic/JuliaGrid.jl}.

\subsection{Determination of Observable Islands}
To evaluate the GBP-based island detection algorithm, we compare the execution time of the GBP-based method $t_{\text{bp}}$ against the numerical method based on the nodal variable formulation $t_{\text{nm}}$ \cite[Sec.~4.6.2]{abur} and the topological method based on the multi-stage procedure $t_{\text{tm}}$ \cite{horisberger}. 

\begin{figure}[ht]
	\centering
	\begin{tikzpicture}
	    \begin{semilogyaxis} [box plot width=0.8mm,
    	    xlabel={Number of Islands per Power System},
        	ylabel={Normalised Time $t_{\text{sa}}/t_{\text{bp}}$},
  	        grid=major,   		
  	        xmin=0, xmax=12, ymax = 700,	
  	        xtick={1,2,3,4,5,6,7,8,9,10,11,12,13},
            xticklabels={,,2-17 islands,,,,,,18-33 islands,,}, 
            xticklabel style={yshift=-10pt},
  	        extra x ticks={1.5,4.5,7.5,10.5,13.5,16.5},
            extra x tick style={xticklabels={1354-bus,2000-bus,1354-bus,2000-bus}, grid=none, major tick length=0pt, ticklabel style={yshift=10pt}},
  	        ytick={0, 1, 10, 100},
  	        width=8cm,height=4.9cm,
  	        tick label style={font=\footnotesize}, label style={font=\footnotesize},
  	        legend style={draw=black,fill=white,legend cell align=left,font=\tiny, at={(0.73,0.62)},anchor=west}]
	
	        \boxplot [
                forget plot, fill=blue!30,
                box plot whisker bottom index=1,
                box plot whisker top index=5,
                box plot box bottom index=2,
                box plot box top index=4,
                box plot median index=3] {./figure/plot1/NumericalTime_2_17_1354.txt};  

	        \boxplot [
                forget plot, fill=red!30, 
                box plot whisker bottom index=1,
                box plot whisker top index=5,
                box plot box bottom index=2,
                box plot box top index=4,
                box plot median index=3] {./figure/plot1/TopologicalTime_2_17_1354.txt};        

	        \boxplot [
                forget plot, fill=blue!30,
                box plot whisker bottom index=1,
                box plot whisker top index=5,
                box plot box bottom index=2,
                box plot box top index=4,
                box plot median index=3] {./figure/plot1/NumericalTime_2_17_2000.txt};   
	        \boxplot [
                forget plot, fill=red!30,
                box plot whisker bottom index=1,
                box plot whisker top index=5,
                box plot box bottom index=2,
                box plot box top index=4,
                box plot median index=3] {./figure/plot1/TopologicalTime_2_17_2000.txt};   
                
	        \boxplot [
                forget plot, fill=blue!30,
                box plot whisker bottom index=1,
                box plot whisker top index=5,
                box plot box bottom index=2,
                box plot box top index=4,
                box plot median index=3] {./figure/plot1/NumericalTime_18_33_1354.txt};  
	        \boxplot [
                forget plot, fill=red!30,
                box plot whisker bottom index=1,
                box plot whisker top index=5,
                box plot box bottom index=2,
                box plot box top index=4,
                box plot median index=3] {./figure/plot1/TopologicalTime_18_33_1354.txt};  
                
	        \boxplot [
                forget plot, fill=blue!30,
                box plot whisker bottom index=1,
                box plot whisker top index=5,
                box plot box bottom index=2,
                box plot box top index=4,
                box plot median index=3] {./figure/plot1/NumericalTime_18_33_2000.txt};  
	        \boxplot [
                forget plot, fill=red!30,
                box plot whisker bottom index=1,
                box plot whisker top index=5,
                box plot box bottom index=2,
                box plot box top index=4,
                box plot median index=3] {./figure/plot1/TopologicalTime_18_33_2000.txt};             
                
            \draw[draw=red, dashed] (5,-1.6) rectangle ++(110,2.75);
            \node[] at (105,0.5) {\footnotesize {$t_{\text{tm}}/t_{\text{bp}}$}};
            \draw[draw=red, dashed] (5,2.5) rectangle ++(110,3.5);
            \node[] at (105,3.1) {\footnotesize {$t_{\text{nm}}/t_{\text{bp}}$}};
	        \draw [thin, dashed] (30,\pgfkeysvalueof{/pgfplots/ymin}) -- (30,\pgfkeysvalueof{/pgfplots/ymax});
	        \draw [thin] (60,\pgfkeysvalueof{/pgfplots/ymin}) -- (60,\pgfkeysvalueof{/pgfplots/ymax});
	        \draw [thin, dashed] (90,\pgfkeysvalueof{/pgfplots/ymin}) -- (90,\pgfkeysvalueof{/pgfplots/ymax});
	        \draw [thin] (120,\pgfkeysvalueof{/pgfplots/ymin}) -- (120,\pgfkeysvalueof{/pgfplots/ymax});
	        \draw [thin, dashed] (150,\pgfkeysvalueof{/pgfplots/ymin}) -- (150,\pgfkeysvalueof{/pgfplots/ymax});
	    \end{semilogyaxis}
	\end{tikzpicture}
	\caption{The normalised execution time of the numerical and topological methods to the GBP-based method for power systems with 1354 and 2000 buses with regard to the number of islands.}
	\label{plot1}
\end{figure}
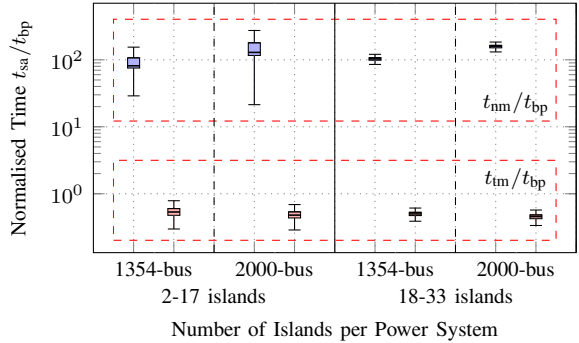

\figurename~\ref{plot1} shows distributions of the normalised execution times $t_{\text{nm}}/t_{\text{bp}}$ and $t_{\text{tm}}/t_{\text{bp}}$ for power systems with 1354 and 2000 buses. We split results into two groups corresponding to the particular range of the number of islands. For all experiments, the topological and GBP-based methods achieve significantly lower execution times compared to the numerical method. Both topological and the GBP-based methods result in comparable execution times across all experiments with the topological method achieving marginally lower execution times. Note that median values of the execution time for the topological and GBP-based methods are between $0.78\,\text{ms}$ and $1.37\,\text{ms}$, and between $1.38\,\text{ms}$ and $2.96\,\text{ms}$, respectively\footnote{The methods have been tested on a 64 bit Windows 10 with an Intel CoreTM i7-8850H 2.60 GHz CPU, and 32 GB of RAM. It is important to emphasise that the same GBP iteration scheme (\ie stopping criterion threshold, the maximum number of iterations) was used throughout all scenarios. However, for the specific system, by tuning these parameters, execution times can be further improved.}.

\begin{figure}[ht]
	\centering
	\begin{tikzpicture}
	    \begin{axis} [box plot width=0.8mm,
    	    xlabel={Number of Islands per Power System},
        	ylabel={Normalised Time $t_{\text{nm}}/t_{\text{bp}}$},
  	        grid=major,   		
  	        xmin=0, xmax=8, ymax = 3,	
  	        xtick={1,2,3,4,5,6,7,8},
            xticklabels={,,}, 
  	        extra x ticks={2,6},
            extra x tick style={xticklabels={2-17 islands,18-33 islands}, grid=none, major tick length=0pt},
  	        ytick={0.5, 1, 1.5, 2, 2.5},
  	        width=8cm,height=4.9cm,
  	        tick label style={font=\footnotesize}, label style={font=\footnotesize},
  	        legend style={draw=black,fill=white,legend cell align=left,font=\tiny, at={(0.01,0.82)},anchor=west}]
	
	       \addlegendimage{area legend,fill=blue!30,draw=black}
	       \addlegendentry{10000-bus}; 
           \addlegendimage{area legend,fill=red!30,draw=black}
	       \addlegendentry{25000-bus}; 
	       \addlegendimage{area legend,fill=green!30,draw=black}
	       \addlegendentry{70000-bus};
	
	        \boxplot [
                forget plot, fill=blue!30,
                box plot whisker bottom index=1,
                box plot whisker top index=5,
                box plot box bottom index=2,
                box plot box top index=4,
                box plot median index=3] {./figure/plot2/TopologicalTime_2_17_10k.txt};  
	        \boxplot [
                forget plot, fill=red!30, 
                box plot whisker bottom index=1,
                box plot whisker top index=5,
                box plot box bottom index=2,
                box plot box top index=4,
                box plot median index=3] {./figure/plot2/TopologicalTime_2_17_25k.txt};  
	        \boxplot [
                forget plot, fill=green!30, 
                box plot whisker bottom index=1,
                box plot whisker top index=5,
                box plot box bottom index=2,
                box plot box top index=4,
                box plot median index=3] {./figure/plot2/TopologicalTime_2_17_70k.txt};                  
	        \boxplot [
                forget plot, fill=blue!30,
                box plot whisker bottom index=1,
                box plot whisker top index=5,
                box plot box bottom index=2,
                box plot box top index=4,
                box plot median index=3] {./figure/plot2/TopologicalTime_18_33_10k.txt};   
	        \boxplot [
                forget plot, fill=red!30,
                box plot whisker bottom index=1,
                box plot whisker top index=5,
                box plot box bottom index=2,
                box plot box top index=4,
                box plot median index=3] {./figure/plot2/TopologicalTime_18_33_25k.txt};   
	        \boxplot [
                forget plot, fill=green!30,
                box plot whisker bottom index=1,
                box plot whisker top index=5,
                box plot box bottom index=2,
                box plot box top index=4,
                box plot median index=3] {./figure/plot2/TopologicalTime_18_33_70k.txt};                 
	        \draw [thin] (400,-20) -- (400,300);
	    \end{axis}
	\end{tikzpicture}
	\caption{The normalised execution time of the topological method versus the GBP-based method for the power systems with 10000, 25000 and 70000 buses with regard to the number of islands.}
	\label{plot2}
\end{figure}
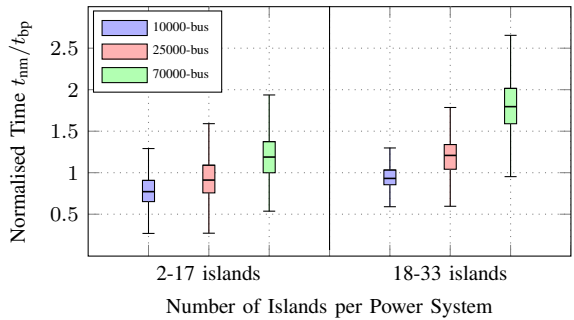

Using a funnel-based approach, we omit numerical method from further experiments due their large execution times on power systems with a smaller number of buses, making it unpractical for larger systems. \figurename~\ref{plot2} shows box plots of the normalised execution time $t_{\text{tm}}/t_{\text{bp}}$ for the power systems with 10000, 25000 and 70000 buses with regard to the number of islands. The GBP-based method \textit{outperforms} topological method for larger power systems. Also, for the fixed power system size (\ie the number of buses), increasing the number of islands results in lower execution times for the GBP-based method compared to the topological method. Note that the execution time medians of the GBP-based method for 70000-buses are $92.97\,\text{ms}$ and $123.43\,\text{ms}$ for 2-17 and 18-33 group of islands, respectively.

\subsection{Observability Restoration}
To evaluate the GBP-based observability restoration algorithm, we compare the execution time of the GBP-based method $t_{\text{bp}}$ against the numerical method based on the Gram matrix associated with a Jacobian matrix corresponding to a reduced network and measurement set $t_{\text{nm}}$ \cite{korres2}. 

\begin{figure}[ht]
	\centering
	\begin{tikzpicture}
	    \begin{axis} [box plot width=0.8mm,
    	    xlabel={Number of Islands per Power System},
        	ylabel={Normalised Time $t_{\text{nm}}/t_{\text{bp}}$},
  	        grid=major,   		
  	        xmin=0, xmax=8, ymin = 0, ymax = 8,	
  	        xtick={1,2,3,4,5,6,7,8},
            xticklabels={,,}, 
  	        extra x ticks={1,3,5,7},
            extra x tick style={xticklabels={2-9 islands,10-17 islands,18-25 islands,26-33 islands}, grid=none, major tick length=0pt},
  	        ytick={1,3,5,7},
  	        width=8.5cm,height=4.9cm,
  	        tick label style={font=\footnotesize}, label style={font=\footnotesize},
  	        legend style={draw=black,fill=white,legend cell align=left,font=\tiny, at={(0.01,0.82)},anchor=west}]

	        \boxplot [
                forget plot, fill=blue!30,
                box plot whisker bottom index=1,
                box plot whisker top index=5,
                box plot box bottom index=2,
                box plot box top index=4,
                box plot median index=3] {./figure/plot3/korres_2_9.txt};  
	        \draw [thin] (200,-20) -- (200,800);
	        
	        \boxplot [
                forget plot, fill=blue!30,
                box plot whisker bottom index=1,
                box plot whisker top index=5,
                box plot box bottom index=2,
                box plot box top index=4,
                box plot median index=3] {./figure/plot3/korres_10_17.txt};  
	        \draw [thin] (400,-20) -- (400,800);
	        
	        \boxplot [
                forget plot, fill=blue!30,
                box plot whisker bottom index=1,
                box plot whisker top index=5,
                box plot box bottom index=2,
                box plot box top index=4,
                box plot median index=3] {./figure/plot3/korres_18_25.txt};  
	        \draw [thin] (600,-20) -- (600,800);
	        
	        \boxplot [
                forget plot, fill=blue!30,
                box plot whisker bottom index=1,
                box plot whisker top index=5,
                box plot box bottom index=2,
                box plot box top index=4,
                box plot median index=3] {./figure/plot3/korres_26_33.txt};  
	    \end{axis}
	\end{tikzpicture}
	\caption{The normalised execution time of the numerical method versus the GBP-based method for 70000-bus power system with regard to the number of islands.}
	\label{plot3}
\end{figure}
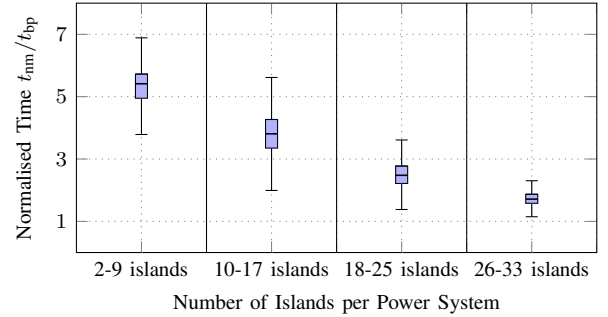

\figurename~\ref{plot3} shows distributions of the normalised execution times $t_{\text{nm}}/t_{\text{bp}}$ for the power system with 70000 buses. We split results into four groups corresponding to the particular range of the number of islands. For all experiments, execution times for both methods are on order of few milliseconds. Still, the GBP-based method achieves lower execution times compared to the numerical method. Median values of the execution time for the GBP-based and numerical methods are between $1.04\,\text{ms}$ and $3.33\,\text{ms}$, and between $5.45\,\text{ms}$ and $5.66\,\text{ms}$, for 2-9 and 26-33 group of islands, respectively. As a sideline, among 100000 random measurement configurations, the numerical method failed to find non-redundant set of equations in $239$ cases due to the incorrect zero pivot threshold, while the GBP-based method succeeded throughout \textit{all} measurement configurations. Note that the redundancy has a significant effect on the number of failed attempts to find the non-redundant set of equations for the numerical method, and increases as the redundancy decreases.
 
\subsection{Computational complexity analysis} 
The complexity of the GBP algorithm, considering only the asymptotic behaviour, depends on: i) the scaling of the number of edges in the factor graph, and ii) the scaling of the number of iterations required for the algorithm to converge.

The first point exploits the underlying system sparsity (i.e., measurement factor nodes are typically connected to a very low number of state variable nodes), as the computational effort per iteration is proportional to the number of edges in the factor graph. For each of the $k$ measurements, the degree of the corresponding factor node is limited by a typically small constant. Indeed, for any type of measurements, the corresponding measurement function depends only on a few state variables. As the number of state variables $n$ and the number of measurements $k$ grows large, the number of edges in the factor graph scales as $\mathcal{O}(n)$. Therefore the computational complexity of the GBP scales \emph{linearly per iteration}. The above analysis is directly related to the sparsity of the underlying factor graph \cite{CosovicTra, zhang}.

For the second point, the scaling of the number of the GBP iterations as $n$ grows large is a more challenging problem related to the convergence rate of the GBP algorithm. Based on a discussion in \cite{bickson} for full matrices, we know that the number of iterations is likely to scale with the condition number of the underlying matrix, which, for well-conditioned matrices may scale as low as $\mathcal{O}(1)$ (constant). In our method, we only track the evolution of the variance values, because variances converge much faster than means. Tracking only variances results in a significant simplification. In our framework, we expect the number of the GBP iterations to scale as a constant.

Another advantage of using the GBP algorithm is the possibility of distributing GBP method across disjoint areas by arbitrarily segmenting the underlying factor graph. In the extreme case of the fully-distributed GBP algorithm, each factor graph node operates locally and independently. In this scenario, the problem is distributed across $\mathcal{O}(n)$ nodes, and, if implemented to run in parallel, it can be $\mathcal{O}(n)$ times faster than the centralised solution.

\section{Conclusions}
We presented a novel computationally efficient GBP-based observability analysis suitable for large-scale data processing in electric power systems that can be easily implemented as part of a fully distributed architecture. In essence, proposed algorithms use only addition operations over the factor graph. Both, determination of observable islands and observability restoration algorithm, are implemented by tracking evolution of the variances, which results in the algorithm that always converges. While algorithms use floating point calculations, most of the time they operate as topological methods. Finally, for large-scale power systems, GBP-based method significantly outperforms numerical and topological state-of-the-art methods, reducing execution time of island detection procedure while being more robust in a search for correct non-redundant set of equations.

\section{Acknowledgment}
This paper has received funding from the European Union's Horizon 2020 research and innovation programme under Grant Agreement number 856967.

\bibliographystyle{IEEEtran}
\bibliography{cite}

\end{document}

%% file: acronyms.tex
\newacronym{se}{SE}{State Estimation}
\newacronym{bp}{BP}{Belief Propagation}

\usepackage{xspace}

\newcommand{\ie}{{i.e.,}\xspace} 
\newcommand{\eg}{{e.g.,}\xspace} 